\documentclass[proceedings, preprint]{rmaa}

\usepackage{paralist}

\usepackage{gensymb}

\SetYear{2020}
\SetConfTitle{Robotic Observatories Workshop}

\title{The Zadko telescope results: ten years of science} 

\author{
  B. Gendre,\altaffilmark{1} \altaffilmark{2} 
  D. Coward,\altaffilmark{1}
  J. Moore,\altaffilmark{1}
	A. Burrell,\altaffilmark{1}
	A. Klotz,\altaffilmark{3}
	P. Thierry,\altaffilmark{4}
	H. Crisp,\altaffilmark{1}
	and E. Howell\altaffilmark{1}}

\altaffiltext{1}{OzGrav-UWA, 35 Stirling Highway, 6009 Crawley, WA, Australia.}
\altaffiltext{2}{E-mail: bruce.gendre@uwa.edu.au}
\altaffiltext{3}{IRAP, 14 avenue Edouard Belin, 31400 Toulouse, France}
\altaffiltext{4}{AGORA observatoire des Makes, 18 Rue Georges Bizet, 97421 La Rivière, France}

\shortauthor{Gendre et al.}
\shorttitle{Zadko Telescope results}

\listofauthors{B. Gendre, D. Coward, J. Moore, A. Burrell, A. Klotz, P. Thierry, H. Crisp, \& E. Howell}
\indexauthor{Gendre, B.}
\indexauthor{Coward, D.}
\indexauthor{Moore, J.}
\indexauthor{Burrell, A.}
\indexauthor{Klotz, A.}
\indexauthor{Thierry, P.}
\indexauthor{Crisp, H.}
\indexauthor{Howell, E.}

\abstract{The 1.0 meter f/4 fast-slew Zadko telescope is located in Western Australia, approximately seventy kilometers north of Perth at Yeal in the Shire of Gingin in a dedicated ”low-luminosity” area. It is the only meter class optical research facility at this southern latitude between the east coast of Australia and South Africa and can rapidly image optical transients at a longitude not monitored by other similar facilities. We review here the main results achieved during the last decade and give some points toward the goals set for future years. Finally we discuss the modifications and improvements we had to perform in the facility to reach these new goals.}

\resumen{El telescopio Zadko de 1,0\,m y focal f$/$4 de giro r\'apido se encuentra en Australia Occidental, aproximadamente setenta kil\'ometros
al norte de Perth en Yeal, en la Comarca de Gingin, en un \'area dedicada de "baja luminosidad". Es la \'unica
instalaci\'on de investigaci\'on \'optica en su clase en esta latitud sur entre la costa este de Australia y Sud\'africa y puede
captar r\'apidamente im\'agenes de transitorios \'opticos en una longitud no monitoreada por otras instalaciones similares. Revisamos aqu\'i los
principales resultados alcanzados durante la \'ultima d\'ecada y damos algunas pìstas hacia las metas establecidas para los pr\'oximos años. Finalmente
discutimos las modificaciones y mejoras que tuvimos que realizar en las instalaciones para alcanzar estos nuevos objetivos.}

\addkeyword{Astronomical Detectors; Observational Astronomy; Telescopes}

\begin{document}
\maketitle

\section{Introduction}
\label{sec:intro}

The Zadko Telescope \citep{cow10} is a 1.0 meter f/4 Ritchey-Chretien telescope located in Western Australia, in the shire of Gingin. It has been constructed by DFM Engineering Ltd, and donated to the University of Western Australia by resource company Claire Energy CEO, J. Zadko. The telescope has been in operation since 1$^{st}$ April 2009. The main properties of the observatory are listed in Table \ref{table1}.

\begin{table}[!bht]\centering
  \setlength{\tabnotewidth}{\columnwidth}
  \tablecols{2}
  \caption{Specifications of the observatory} \label{table1}
  \begin{tabular}{lr}
    \toprule
    Property & Value \\
    \midrule
		Observatory code    & D20 \\
		Longitude           & 115\degree 42' 49'' E \\
		Latitude            & 31\degree 21' 24'' S \\
		Altitude            & 50 meters \\
		Seeing              & Variable (1-4\arcsec) \\
		Limiting magnitude  & R$\sim$21.5 (180s exposure) \\
		Camera              & QHY 160M \\
		Filters             & Sloan g', r', i', Clear \\
    \bottomrule
  \end{tabular}
\end{table}

In mid 2010, the telescope control system was fully upgraded using the TAROT robotic telescope software system \citep{klo08}. This suite of independently running programs comprises two main components: AudeLA\footnote{\url{http://www.audela.org/english_audela.php}} and ROS (Robotic Observatory Software). There is also a third party software (mostly drivers) for the mechanical and ancillary systems of the observatory (i.e. the telescope and the weather station). All technical details are reported in \citet{moo20}.

The key feature of the telescope is the optimization for performing automated follow-up from alerts or triggers from different external instruments. In addition, it employs a dynamic scheduling system that allows other projects to operate when not in alert mode. This sets the facility apart from many other systems that are ``internet'' telescopes that still require some level of manual monitoring. Furthermore the location of the facility at a longitude not covered by many other meter class facilities provides an important resource for many projects. 

The scientific goal of the Zadko telescope is to explore a previously uncharted region of the `transient sky' which includes: 
\begin{itemize} 

\item Gamma ray bursts: Rapid optical follow-up of {\it Swift} Alerts.

\item Antares neutrino alert follow-up.

\item Solar system studies: follow-up of asteroids and studies of ``Barbarian asteroids''.

\item Gravitational wave follow-up of events detected by LIGO-Virgo.

\item Nearby supernovae and lensed supernovae search.

\item Education outreach and training.

\item Emerging projects including automated follow-up of radio transients and space debris tracking.

\end{itemize} 

The Zadko Telescope having fulfilled its purpose over a ten year period and producing many publications was shut down in late 2018 for a long overdue maintenance. This included shipping both Primary and Secondary mirrors back to the USA for stripping and re-aluminization \citep[see][for details]{moo20}. 

In this paper, we briefly review the results obtained with the telescope on gamma-ray bursts (section \ref{grb}), and gravitational waves (Section \ref{gw}) taken before the shut down, and discuss the future of the instrumentation (Section \ref{future}).


\section{Gamma-ray bursts}
\label{grb}
We are using a similar strategy to the TAROT network to perform the follow-up of GRBs detected by high energy instruments. In short, the telescope first performs a trailed observation for 30 seconds, allowing a temporal resolution of about 1 second of the prompt light curve, continuing with observations using the clear filter for the next hour.  After that time, where the temporal coverage of the afterglow can allow for discontinuous observations, we are alternating filters to gather color information. The observations are repeated the next night in case of bright events.

Most of the results have been reported in \citet{cow17}, and only two events have not been published yet: GRB 160127A (Crisp et al., in preparation), and GRB 170202A (Gendre et al., in preparation). During nine years of activity the Zadko telescope performed 39 follow-ups of GRBs and detected 13 of them. This rate is common for a robotic telescope placed in a remote location.

\section{Gravitational waves}
\label{gw}
The scientific interest of the telescope has slightly shifted from gamma-ray bursts to the counterparts of gravitational wave emitters with the detections of the first events by the LIGO-Virgo collaboration \citep{abb16, abb17}. The Zadko telescope has been part of the follow-up campaign of GW170817, and has provided very important optical data \citep{and17}. Its limited field of view, however, prevents us to realistically find the counterpart inside a large position error box. The telescope entered the GRANDMA collaboration \citep{ant20} for that purpose. The GRANDMA collaboration aims at coordinating observations between telescopes with a large field of view (to detect potential counterparts) and ones with a small field of view (to confirm/reject the association). Since the start of the run O3, the Zadko telescope participated in three observation campaigns of LIGO-Virgo events. These observations are reported in \citet{ant20}.

\section{Future of the instrumentation}
\label{future}
Maintenance of the telescope has allowed for a pause in the observations and provided time to reflect on the scientific goals. In addition, the experience gained during the first follow-up of fast radio-bursts and gravitational wave events has shown that an evolution of the telescope instrumentation was required. Initially we replaced the previous camera (being an Andor Ikon-L with a CCD) with a new QHY camera equipped with a CMOS chip. We chose a low price camera with a small field of view in order to test this new technology and to be sure it would face the challenges of our scientific goals. Having now confirmed that this technology is suitable for our needs, we are seeking to adapt a large CMOS camera behind the telescope.

We have also confirmed that we require a larger choice of filters to fulfill all our requirements, thus modifying the whole optical path due to limitations with our existing filter wheel. This led us to redesign the mechanical link between the instrument and the telescope enabling an increase in the number of instruments one could mount on the telescope. Lastly, we are investigating the possibility to have a full field low resolution spectrometer, which would allow us to discriminate the various kinds of transients when they are detected, and not during a later follow-up.

It is expected that with all these new modifications, and the mechanical refurbishment of the telescope discussed in \citet{moo20}, the Zadko telescope will be able to fulfill its scientific goals for another ten fruitful years.

\end{document}